\documentclass[prl,aps,twocolumn,amsmath,amsfonts,showpacs,superscriptaddress,preprintnumbers,nofootinbib]{revtex4-2}

\usepackage{graphicx}
\usepackage{xcolor}
\usepackage{rotating}
\usepackage{bm,amsmath,amssymb}
\usepackage[utf8]{inputenc}

\usepackage[colorlinks=true, urlcolor=blue, linkcolor=blue, citecolor=blue, hyperindex=true, linktocpage=true]{hyperref}

\def\be{\begin{equation}}
\def\ee{\end{equation}}

\def\bea{\begin{eqnarray}}
\def\eea{\end{eqnarray}}

\begin{document}
%\preprint{}

%\title{Real-time response of quantum perfect fluids}

\title{Quantum dynamics of perfect fluids}

\author{Walter D. Goldberger}
\thanks{Email: walter.goldberger@yale.edu}
\affiliation{Department of Physics, Yale University, New Haven, CT 06520, USA}
\author{Petar Tadi\' c}
\thanks{Email: petar.tadic@physics.ox.ac.uk}
\affiliation{Department of Physics, University of Oxford, Oxford, OX1 3PU, United Kingdom}
\affiliation{Institute for Interdisciplinary and Multidisciplinary Studies, University of Montenegro, Podgorica, Montenegro}

\begin{abstract}
\noindent We study the quantum field theory of zero temperature perfect fluids.   Such systems are defined by quantizing a classical field theory of scalar fields $\phi^I$ that act as Lagrange coordinates on an internal spatial manifold of fluid configurations. Invariance under volume preserving diffeomorphisms acting on these scalars implies that the long-wavelength spectrum contains vortex (transverse modes) with an exact $\omega_T=0$ dispersion relation.  As a consequence, physically interpreting the results obtained via perturbative quantization of this theory has proven to be challenging.   In this paper, we show that correlators evaluated in a  class of semi-classical (Gaussian) initial states prepared at $t=0$ are well-defined and accessible via perturbation theory. The width of the initial state effectively acts as an infrared regulator without explicitly breaking diffeomorphism invariance of the classical action. As an application, we compute the  stress tensor two-point correlators and show that vortex modes give a non-trivial contribution to the response function,  non-local in both space and time.

\end{abstract}
\maketitle

\section{Introduction}
\noindent Attempts to find a sensible physical interpretation of the quantum theory of zero temperature \emph{ordinary} perfect fluids, i.e., those in which the velocity operator possesses both irrotational and divergence free (vortex) components, have a long history.   The earliest proposal, dating back to the work of Landau~\cite{Landau:1941lul}, is that quantum effects somehow generate a mass gap for the vortex modes, so that the long wavelength dynamics of this system are indistinguishable from those of a superfluid, in which only longitudinal phonon excitations propagate.

In recent years, this problem was revisited in ref.~\cite{Endlich:2010hf}, which studied the problem from the point of view of modern quantum field theory, starting from an effective field theory (EFT) re-formulation of ideal classical hydrodynamics developed in~\cite{Dubovsky:2005xd,Dubovsky:2011sj} (see also~\cite{10.1098/rspa.1972.0164,Carter1973,Soper2008CFT} for earlier work).  In this approach, the (relativistic) perfect fluid is viewed as a non-linear sigma model of scalar fields $\phi^I(x)$, $I=1,2,\cdots,d$ which are maps from $d+1$-dimensional spacetime into a spatial internal manifold of fluid configurations.  The defining property of the perfect fluid is that, in addition to having spacetime Poincare invariance,  the action is also invariant under volume preserving re-parameterizations (``SDiffs'') acting on the Lagrange coordinates, 
\begin{eqnarray}
\phi^I\rightarrow {\hat\phi}^I={\hat\phi^{I}}(\phi), & \det\left({\partial{\hat\phi}^I\over\partial\phi^J}\right)=1.
\end{eqnarray}

A given classical field configuration $\langle\phi^I(x)\rangle\neq 0$ (i.e. a solution of the classical hydrodynamic equations)  spontaneously breaks the combined spacetime and internal diffeomorphisms down to a subgroup.   For example, a non-gravitating static fluid in its rest frame, with four-velocity $u^\mu=\delta^\mu{}_0$, corresponds to a background $e^I_\mu=\langle \partial_\mu \phi^I\rangle = \delta^I_\mu$, which breaks both the spacetime and the internal symmetries, but preserves residual rigid spatial translations and rotations acting simultaneously on the spacetime and internal fluid coordinates.   The pattern of spontaneous symmetry breaking then results in a set of $d$ Goldstone modes $\pi^i(x),$ $i=1,\cdots,d,$  that describe the propagation of both longitudinal (``phonon’’) and transverse (``vortex’’) sound waves about the static background.   The self-interactions of these modes are dictated by the underlying spacetime and internal diffeomorphism invariance of the fluid action and at long wavelengths are suppressed by at least two spacetime derivatives.

From the point of view of the EFT of~\cite{Dubovsky:2005xd,Dubovsky:2011sj}, the difficulties that arise in attempting to quantize the perfect fluid are already apparent in the extreme long wavelength limit in which the Goldstone mode self-interactions can be neglected.    In the free theory, the longitudinal waves have the familiar dispersion relation $\omega_L({\vec k}) = c_s |{\vec k}|$, so that upon quantization the theory possesses a standard Fock spectrum of multi-particle phonon asymptotic states.  However, because the vortex modes, with $\nabla\cdot{\vec\pi}=0$, are themselves generators of infinitesimal SDiffs, the underlying symmetry requires that they possess an exact dispersion relation $\omega_T({\vec k})=0$ that is independent of wavenumber ${\vec k}$, leading to an infinitely degenerate spectrum of zero-energy excitations.   Consequently there is no stable or even normalizable vacuum state.   Given this situation, it is unclear to what extent it is possible to define sensible observables, let alone make predictions for how they evolve in time.

A practical way of handling the infinite degeneracy of vortex zero modes is to deform the theory by suitably small SDiff breaking terms, such that the classical vortex waves acquire a dispersion relation $\omega_T({\vec k})=c_T |{\vec k}|$, where the small parameter $c_T\ll 1 $ characterizes the amount of explicit symmetry breaking.   For $c_T\neq 0$, the quantum theory has a well-defined spectrum of asymptotic multi-particle states, and it is possible to calculate $S$-matrix elements of low-energy modes  perturbatively in a gradient expansion, using standard EFT techniques.   One then attempts to construct out of these amplitudes a set of ``IR safe’’ observables which have a smooth $c_T\rightarrow 0$ limit.    

Following this procedure, ref.~\cite{Endlich:2010hf} calculated tree-level  $3$- and $4$-point amplitudes involving (for $c_T\neq 0$) asymptotic $L$ and $T$ states.    The natural tree-level observables (i.e. scattering cross sections and decay rates) were found to contain phase space singularities that blow up as inverse powers of $c_T$ in the $c_T\rightarrow 0$ limit. Because in the EFT there are no radiative (loop) corrections arising at the same order in the derivative expansion, it is not possible to identify any clear inclusive scattering observables that are free of singularities as the vortex speed of sound vanishes.  The lack of well-defined $c_T\rightarrow 0$ tree-level observables then could be interpreted as a hint of why  perfect fluid phases of bulk matter have not been observed in low temperature experiments, i.e. as a speculative answer to the question of why in nature the $T=0$  ground states of many-body systems seem to only exhibit solid or superfluid phases.

We have verified by an explicit calculation that  $c_T\rightarrow 0$ power divergences similar to those seen at tree-level  also appear in the energy-momentum tensor's two-point connected Wightman correlator at one-loop order (subleading order in the derivative expansion).     In hindsight, it is clear that at $c_T\neq 0$ such singularities must be present in Green's functions of generic local operators, since quantum mechanical unitarity requires the one-loop diagrams to have cuts corresponding to tree-level decay processes.  Even worse, we find by power-counting that multi-loop corrections to correlators scale as successively more powers of $1/c_T$.  This again follows from unitarity, and reflects the growing dimensionality (and therefore volume) of phase space as more particles (cut propagators) are included in the final state.    This pattern indicates a breakdown of the perturbative expansion of the regularized EFT, and it is an open problem to determine if the strongly-coupled nature of the $c_T\rightarrow 0$ limit can be somehow tamed by a judicious resummation of the naive loop expansion\footnote{The authors of~\cite{Gripaios:2014yha} claim instead that time-ordered correlators of SDiff invariant operators constructed out of the field $\phi^I(x)$ are IR-safe, at least in $d=2$ spatial dimensions, and can therefore be calculated even without the addition of symmetry breaking deformations to the perfect fluid Lagrangian. However, SDiff invariance of observables is clearly not sufficient to ensure IR safety.   For example, the regularized free theory exhibits pathological thermodynamic behavior, e.g., a vortex mode contribution to the thermal entropy $\sim T^d/c_T^d$ at finite temperature $T$ that diverges in the $c_T\rightarrow 0$ limit.    We find that even in the context of the Green's functions considered in~\cite{Gripaios:2014yha}, the claims of IR safety are an artifact of the specific form of dimensional regularization (in which the spatial and time directions have to be separately continued away from the physical dimensions), and, as we have verified, do not hold if other regularization schemes are used to define the one-loop integrals.  See Appendix~\ref{app:triumph} for an explicit comparison of results using different regulators.}

Given the difficulties of making sense of the perturbative quantization of the theory, ref.~\cite{Dersy:2022kjd} pursued instead an algebraic approach, in which the spectrum is constructed non-perturbatively, from the representation theory of the algebra of SDiff's acting on the sigma model target space (taken to be a 2D torus $T^2$ in~\cite{Dersy:2022kjd}).   

This line of analysis leads to interesting and calculable observables.  Among the robust predictions of this approach is the identification of a normalizable vacuum state and a well-defined Fock spectrum of particles with a non-relativistic $E={\vec p}^2/2M$ dispersion relation and a perturbative $S$-matrix mediated by a quartic spatially non-local interaction (the $\mbox{SDiff}(T^2)$ representation theory allows these particles, the ``vortons’’, to be realized as either fermions or bosons).   The physical mass scale $M$  that regularizes the IR physics of this system arises radiatively, due to one-loop corrections generated by high-momentum virtual particles, and has the effect of spontaneously breaking the full $\mbox{SDiff}(T^2)$ completely in the vorton vacuum~\cite{Cuomo:2024ekf}.   Ref.~\cite{Cuomo:2024ekf} also argued, by employing a Clebsch parametrization of the fluid variables, that the vorton picture does not rely on special features of hydrodynamics in 2D, applying as well to 3D fluids.   See also ref.~\cite{Tsaloukidis:2025jzj} which provides an interpretation of the non-perturbative 2D vortex modes as topological defects, and introduces variables directly analogous to the ones used to describe dislocations in elastic media. Other attempts to regularize the EFT include refs.~\cite{Torrieri:2011ne,Burch:2015mea,Montenegro:2020paq}. 
Note also alternative formulations of perfect fluid EFT incorporating quantum effects in flows with nonzero vorticity~\cite{Wiegmann:2019mdg}.

While the results established in~\cite{Dersy:2022kjd,Cuomo:2024ekf} address the stability of the quantum theory and its excitations near the vacuum, they do not make contact with the phenomena usually associated with hydrodynamics, namely macroscopic flows and the corresponding transport of conserved quantities.   In this paper, we study these dynamical properties of quantum hydrodynamics in initial states that are far from the vacuum, for which the original parametrization of the perfect fluid EFT, written in  terms of the Lagrange variable $\phi^I(x)$ is better suited.  By focusing on a specific example of the stress-tensor response function, we show that it is possible after all to make well-defined predictions for this observable which do not depend on any ad-hoc regulators introduced to artificially lift the infinite degeneracy of vortex modes in the IR. 

Our calculation is based on the observation, already emphasized in~\cite{Endlich:2010hf}, that in the free limit, the $\omega_T({\vec k})=0$ dispersion relation  of the vortex modes (linear in time growth of fluctuations) implies that their free quantum dynamics is analogous to that of an infinite collection of non-relativistic free particles, indexed by a spatial momentum label.   Consequently, as in non-relativistic quantum mechanics of free particles, the stationary states are non-normalizable and therefore live outside the Hilbert space of physical states.  Rather, the physical states are normalizable wavefunctions (localized wavepackets) prepared by an experimentalist on some initial time slice.   As these states evolve, their wavefunctions spread out and eventually, at $t\rightarrow\infty$ completely delocalize.   

This behavior then accounts for some of the pathologies encountered in perturbative calculations as $c_T\rightarrow 0$.   For finite $c_T$, correlators are implicitly defined (via an $i\epsilon$ prescription) to be taken in the $c_T\neq 0$ ground state of the $T$ modes, a Gaussian wavefunctional $\Psi_{-\infty}[\phi^I]$ at $t=-\infty$.   By the time local operators on finite spatial slices are inserted into the functional integral, this Gaussian state would evolve in the $c_T\rightarrow 0$ limit into the zero state  $\Psi_{t}[\phi^I]\equiv 0$ for any finite value of time $t$, rendering even the perturbative path integral incalculable.      In contrast, if we instead start with a normalizable state defined at some finite time ($t=0$) and restrict ourselves to operator insertions at sufficiently short times $t>0$, then the path integral is calculable in perturbation theory, with predictions that depend only on (in principle) experimentally measurable properties of the state prepared at $t=0$.  For initial states that are \emph{not} SDiff invariant, the insertion of the wavefunction into the Schwinger-Keldysh path integral renders the $T$ mode propagators non-degenerate, without the need to add IR regulator terms to the effective Lagrangian which might spoil any of the symmetries that define the classical perfect fluid theory.

To illustrate these remarks, we will focus in this paper on the retarded response function of the perfect fluid stress tensor $T_{ij}(x),$
\begin{equation}\nonumber
G^{ij,kl}_R(x_1,x_2) = -i\theta(x_1^0-x_2^0)\langle \psi_i|[T^{ij}(x_1),T^{kl}(x_2)]|\psi_i\rangle,
\end{equation}
evaluated in an initial state  $|\psi_i\rangle$ at $t=0$ belonging to a class of semi-classical wavefunctions that we discuss in more detail below.  We show by an explicit one-loop calculation that, for a suitable initial state and at least for sufficiently short times, this is given by a well-defined function that (i) depends on contributions from both phonon and vortex modes (i.e., it differs from the response of a pure superfluid), (ii) is free of UV and IR divergences, and (iii) depends only on (in principle) experimentally measurable properties of the perfect fluid at the initial time  $t=0$ where the experiment starts.

\section{General formalism}
\noindent In order to calculate Green's functions, we start with the Schwinger-Keldysh (in-in) representation of correlation functions in a given initial state $|\psi_i\rangle$ at $t=0$.  If $T[{\cal O}_1[\phi^I]]$ (${\tilde T}[{\cal O}_2[\phi^I]]$) denotes the time-ordered (anti time-ordered) product of a string of generic local operators, the in-in representation of operator expectation values in the state $|\psi_i\rangle$ is given by the formal path integral expression 
\begin{widetext}
\begin{equation}
\label{eq:inin}
\langle \psi_i|{\tilde T}[{\cal O}_2[\phi^I]] T[{\cal O}_1[\phi^I]]|\psi_i\rangle = \int^{\phi^I_+({\vec x},\tau)=\phi^I_-({\vec x},\tau)} {\cal D}\phi_-^I(x) {\cal D}\phi_+^I(x)\Psi_i[\phi^I_-]^* \Psi_i[\phi^I_+] e^{i S[\phi^I_+]-iS[\phi^I_-]}  {\cal O}_1[\phi^{I}_+] {\cal O}_2[\phi^{I}_-],
\end{equation}
\end{widetext}
where future boundary conditions, with the fields on either branch of the closed-time contour identified, are imposed at an arbitrary time slice $t=\tau$ taken to occur later than any operator insertions inside the functional integral\footnote{ Ultimately, quantum mechanical unitarity guarantees that any dependence on $\tau$ must drop out of physical quantities computed using the in-in path integral.  We have explicitly verified that such cancellations take place in tree level two- and three-point functions, independently of the specific form of the interactions.}.  In the above expression, $\Psi_i[\phi^I({\vec x})]=\langle \phi^I|\psi_i\rangle$ is the Schrodinger functional representation of the initial state $|\psi_i\rangle$ expanded in the basis of eigenstates of the field operator $\phi^I({\vec x})$ at $t=0$.  See~\cite{Calzetta:2008iqa} for a pedagogical introduction to Schwinger-Keldysh methods in quantum field theory.

 The classical action $S[\phi^I]$ of the fluid is constrained by symmetry to be an arbitrary local functional of SDiff invariants constructed from the coordinates $\phi^I(x)$.   In the leading (two-derivative) long-wavelength limit,  such invariants consist of arbitrary real analytic functions of the determinant $B=\left|\det B^{IJ}\right|$ of the (inverse) metric $B^{IJ}=\eta^{\mu\nu} \partial_\mu \phi^I \partial_\nu \phi^J$ on Lagrange space.  Thus, at low energies and momenta, $S[\phi^I]$ takes the form
 \begin{equation}
\label{eq:action}
S[\phi^I]=w_0\int d^{d+1} x f(\sqrt{B}),
\end{equation}
where the dimensionless function $f(\sqrt{B})$, which we normalize to $f'(1)=1$, defines the equation of state $p=p(\rho)$ of the perfect fluid.   Comparing to the perfect-fluid $T^{\mu\nu}=(\rho+p) u^\mu u^\nu - p \eta^{\mu\nu}$, one finds in particular that the comoving energy density is $\rho(x)=-w_0 f(\sqrt{B})$.    Because the Lagrange coordinate $\phi^I$ is comoving, i.e., invariant along the fluid flow, the components of the fluid's four-velocity field $u^\mu(x)$ satisfy the constraints $u^\mu\partial_\mu \phi^I(x)=0$.   This constraint determine the three-velocity in any fixed Lorentz frame to be
\be
v^i(x) = - {\partial x^i\over \partial \phi^I} \partial_t \phi^I.
\ee

In our calculation below, we consider an initial state at $t=0$ that most closely resembles what one would regard as a classical fluid configuration, i.e., one in which the expectation value $\langle \phi^I({\vec x})\rangle\neq 0$ at $t=0$ spontaneously breaks Poincare$\times$SDiff invariance, and in which the product of uncertainties in the initial value of $\phi^I({\vec x})$ and its conjugate momentum $\Pi_I=\partial {\cal L}/{\partial{\dot \phi}^I}$ is as small as possible.   These requirements suggest that we take $\Psi_i[\phi^I]$ to be a minimum uncertainty wavepacket centered at the VEV $\langle \phi^I({\vec x})\rangle\neq 0$, i.e. a Gaussian functional of the general form
{ 
\bea\nonumber
\label{eq:state}
\Psi_i[\phi^I]&={\cal N}&\exp\left[-{1\over 2}\int d^d {\vec x} d^d{\vec y} \pi^I({\vec x}) K_{IJ}({\vec x},{\vec y})\pi^J({\vec x})\right]\times\\
&  &\exp\left[ i \int  d^d{\vec x} v_I({\vec x}) \pi^I({\vec x}) \right]
\eea
where $\pi^I({\vec x})=\phi^I({\vec x})-\langle \phi^I({\vec x})\rangle$, and ${\cal N}$ is chosen such that $|\psi_i\rangle$ is properly normalized.
}

Even though the functions $K_{IJ}({\vec x},{\vec y})$ and $v_I({\vec x})$, can be chosen fairly arbitrarily, they have direct physical interpretation in terms of the expectation values of fluid observables $t=0$.     In particular, the expectation value of a generic operator  ${\hat O}[\phi^I,\Pi_J]$ at $t=0$ (where $\Pi_I({\vec x})=-i\delta/\delta\phi^I({\vec x})$ in the Schrodinger functional picture) is determined by a Gaussian path integral of the form
\be
\langle \psi_i| {\hat O}|\psi_i\rangle = {\int {\cal D\phi}^I({\vec x})\Psi^*_i[\phi] {O}[\phi^I,\Pi_J]\Psi_i[\phi]\over \int {\cal D\phi}^I({\vec x}) |\Psi_i[\phi]|^2}.
\ee
For an initial VEV $\langle \phi^I({\vec x})\rangle\neq 0$, we define ${\bar e}^I_i({\vec x}) = \langle \partial_i \phi^I({\vec x})\rangle$, and denote its inverse matrix by ${\bar e}^i_I({\vec x})$ and determinant ${\bar e}=\det {\bar e}^I_i$. Then to linear order in the fluctuations $\pi^I$, the velocity operator at $t=0$ is
\be
v_i({\vec x}) = -{i{\bar e}^I_i({\vec x}) \over w_0{\bar e}f'({\bar e})} {\delta\over\delta \pi^I({\vec x})}+\cdots, 
\ee 
while the density fluctuation is 
\be
\delta\rho({\vec x})={\rho}({\vec x})-{\bar \rho}({\vec x}) = -{ w_0 f'({\bar e}) \over 2{\bar e}} {\bar e}^i_I \partial_i \pi^I({\vec x})+\cdots.
\ee
It follows that in our initial state $\Psi_i[\phi^I]$, the velocity has $t=0$ expectation value $\langle v_i({\vec x})\rangle = {\bar e^I_i v_I({\vec x})/w_0 {\bar e}f'({\bar e})}$, while velocity fluctuations $\delta v_i =v_i - \langle v_i\rangle$ have variance directly proportional to the kernel $K_{IJ}$ in Eq.~(\ref{eq:state}),
\bea
\nonumber
\langle \delta v_i({\vec x}) \delta v_j ({\vec y}) \rangle = {1\over 2}  \left[{{\bar e}^I_i  ({\vec x}) \over w_0 {\bar e}f'({\bar e})}\right]K_{IJ}({\vec x},{\vec y}) \left[{{\bar e}^J_j ({\vec y}) \over w_0 {\bar e}f'({\bar e})}\right].\\ 
\eea
Similarly, the $t=0$ density-density correlator is directly proportional to the \emph{inverse} kernel $[K^{-1}]^{IJ}({\vec x},{\vec y})$ (related to $K_{IJ}({\vec x},{\vec y})$ in the obvious way),
\bea
\nonumber
\langle \delta \rho({\vec x}) \delta \rho ({\vec y}) \rangle &=& \left[{w_0f'({\bar e}) \over 2{\bar e}} {\bar e}^i_I({\vec x})\right] \left[{w_0f'({\bar e}) \over 2{\bar e}} {\bar e}^j_J({\vec y})\right]\\
& & {}  \hspace{0.5cm} \times{1\over 2}\partial^{\vec x}_i  \partial^{\vec y}_j [K^{-1}]^{IJ}({\vec x},{\vec y}).
\eea
In this way, the properties of our chosen initial state are fully fixed in terms of one- and two-point correlators that can be (in principle) determined experimentally.  (Note that even though the initial state is Gaussian, SDiff invariant operators such as the fluid 3-velocity or energy density are in general non-linear functions of $\pi^I({\vec x})$ and therefore such observables exhibit non-Gaussianities already at $t=0$.).

For the remainder of this paper, we will focus on the case of an initial state that closely corresponds to the static, homogeneous and isotropic configuration of a classical space-filling fluid in its rest frame.   To that end, we take $v_I({\vec x})=0$, $\langle\partial_i\phi^I({\vec x})\rangle= \delta^I_i$ (and therefore ignore the distinction between Lagrange $(I)$ and Euler $(i)$ spatial indices from now on) and choose a translation invariant wavefunction kernel $K_{IJ}({\vec x},{\vec y})=K_{ij}({\vec x}-{\vec y})$.   Rotational invariance implies that in generic spatial dimension $d$ the momentum space kernel $K_{ij}({\vec p})=\int d^d {\vec x} e^{i {\vec p}\cdot {\vec x}} K_{ij}({\vec x})$  has the form
\be
\label{eq:generic}
K_{ij}({\vec p}) =P^L_{ij}({\vec p}) K_L({\vec p}) +P^T_{ij}({\vec p}) K_T({\vec p}),
\ee 
with $P_{ij}^L({\vec p})=p_i p_j/{\vec p}{\,}^2$ and $P^T_{ij}({\vec p})=\delta_{ij}-P^L_{ij}$ the longitudinal and transverse projectors respectively, and associated kernels $K_{L,T}({\vec p})$ that depend only on $|{\vec p}|$.

We note, however, that in the cases $d=2,3$ of primary physical interest, it is possible to include parity violating terms as well.  In $d=3$, the general $SO(3)$ invariant kernel takes the form
\bea
\nonumber
K^{d=3}_{ij}({\vec p}) &=&P^L_{ij}({\vec p}) K_L({\vec p}) +P^T_{ij}({\vec p}) K_T({\vec p})\\
& & {}+ i \epsilon_{ijk} {p^k\over |{\vec p}|} K_A\left({\vec p}\right),
\eea
which, for $K^A({\vec p})\neq 0$, allows independent control over the probability distributions of plus and minus helicity vortex quantum fluctuations in the initial state.  In the case of $d=2$ spatial dimensions, there can even be `mixing' between $L$ and $T$ modes induced by parity violating terms in the wavefunction, 
\bea
\nonumber
K^{d=2}_{ij}({\vec p})& &=P^L_{ij}({\vec p}) K_L({\vec p})  + P^T_{ij}({\vec p}) K_T({\vec p})\\
& & + {1\over 2} \left(\epsilon_{ik} P^L_{kj}({\vec p})+\epsilon_{jk} P^L_{ki}({\vec p}) \right) K_{LT}({\vec p}).
\eea 
Here we will limit ourselves to the parity conserving case, $K_A({\vec p})=K_{LT}({\vec p})=0$, leaving a more systematic study of parity violating contributions to $t>0$ response functions for future work.  

For the longitudinal mode wavefunction, we take
\be
K_L({\vec p}) =  w_0 c_s |{\vec p}|,
\ee
which corresponds to the standard free phonon Fock vacuum, with $t=0$ two-point correlators of the density and of the irrotational velocity component that both scale with distance as $\langle \delta\rho({\vec x})\delta\rho(0)\rangle\sim\langle v^L_i({\vec x})v^L_j(0)\rangle\sim 1/|{\vec x}|^{d+1}$.

The choice of transverse kernel $K_T({\vec p})$ is less constrained, but as a representative example for the purpose of this paper, we consider a power-law function of the form 
\be
K_T({\vec p}) =  w_0 {\hat c}_T\mu \left({|{\vec p}|\over \mu}\right)^{\Delta},
\ee
with free parameters $\mu>0$, which sets the overall momentum scale, spectral index $\Delta$, and ``transverse speed of sound'' ${\hat c}_T>0$.  As long as 
\be
\Delta>-d, 
\ee
the kernel suppresses high-momentum fluctuations, while at the same time ensuring that equal-time correlators at $t=0$ exhibit power-law spatial correlations that decay at long distances, with $t=0$ transverse (divergence free) velocity correlator $\langle v^T_i({\vec x})v^T_j(0)\rangle\sim 1/|{\vec x}|^{d+\Delta}$.   

Note that for the choice $\Delta =1$, the wavefunction of the transverse modes is formally identical to the Fock vacuum of free bosons with non-zero speed of sound ${\hat c}_T\neq 0$.  However, we emphasize that the parameter ${\hat c}_T$ has a different interpretation than the IR regulator $c_T\neq 0$ introduced in~\cite{Endlich:2010hf} to make sense of the perturbative expansion.   The latter corresponds to the coefficient of the Lagrangian deformation $\Delta {\cal L} = -{1\over 2} c_T^2 w_0 \delta_{IJ}B^{IJ}$, which explicitly breaks SDiffs at the level of the dynamics, while the former only breaks the symmetry at $t=0$ without modifying the classical fluid Lagrangian Eq.~(\ref{eq:action}) for $t>0$.  It  serves to define the characteristic energy scale $\mu {\hat c}_T$ of vortex modes in the initial state at $t=0$, and therefore is not constrained by relativistic causality to be bounded above by unity.

Having fixed the form of the fluid action and the initial state wavefunction, it is now possible to calculate matrix elements in the state $\Psi_i$ perturbatively, by changing variables to $\pi^I_{\pm}(x)= \phi^I_\pm(x)-\langle \phi^I\rangle$ in Eq.~(\ref{eq:inin}) and expanding in powers of the fluctuation.  This gives rise to a set of well-defined Feynman rules with propagators and interaction vertices that can be read off Eq.~(\ref{eq:inin}).  In particular, the (Wightman) propagators follow from the quadratic part of the fluid Lagrangian
\be
\mathcal{L}=w_0\left(\frac{1}{2}\dot{\vec{\pi}}^{2}-\frac{1}{2}c_{s}^2 (\nabla\cdot{\vec\pi})^2+\cdots\right), 
\ee
with $c_s^2 =f''(1)$, together with the wavefunction insertions in Eq.~(\ref{eq:inin}).  Given the form of the wavefunction in Eq.~(\ref{eq:generic}), it is natural to formulate the Feynman diagram expansion in a mixed time-momentum representation $(t,{\vec p})$, in which case the momentum space Wightman function $W_{\vec p}^{ij}(x^0,y^0)=\int d^d {\vec x} e^{i{\vec p}\cdot({\vec x}-{\vec y})} \langle \pi^i(x)  \pi^i(y)\rangle_0$ of the free field has a decomposition into separate $L$ and $T$ sectors
\be\label{eq:Wightman2pt}
W_{\vec p;ij}(x^0,y^0)= P^L_{ij}({\vec p}) W_{\vec p}^{L}(x^0,y^0)+P^T_{ij}({\vec p})  W_{\vec p}^{T}(x^0,y^0).
\ee
For the choice $K_L({\vec p}) = w_0 c_s |{\vec p}|$, we have, from Eqs.~(\ref{eq:inin}),~(\ref{eq:state}),~(\ref{eq:generic}),
\be
W_{\vec p}^{L}(x^0,y^0)= {e^{-i c_s |{\vec p}| (x^0-y^0)}\over 2 w_0 c_s |{\vec p}|},
\ee
while
\be
W_{\vec p}^{T}(x^0,y^0)= {\left[1 - {i x^0\over w_0} K_T(|{\vec p}|)\right]\left[1 + {i y^0\over w_0} K_T(|{\vec p}|)\right]\over 2 K_T({\vec p})},
\ee
for a general choice of transverse kernel $K_T({\vec p})$.  Notice that despite the absence of a vortex mode energy gradient term in the Lagrangian, the insertion of the initial state wavefunction at $t=0$ ensures that its propagator is non-degenerate, at least if observed at finite times $x^0,y^0>0$.  

In general, the expansion in Feynman diagrams will be under control when both external momenta and times are sufficiently small.  Specifically, the power counting of the EFT is such that 
\begin{itemize}
\item \emph{Derivative expansion}: In order for loop/derivative corrections to be suppressed, spatial momenta and time derivatives must be small, $|{\vec p}|^{d+1}/w_0\ll1$ or $|{\vec p}|^{d-1} \partial_t^2 /w_0\ll 1$.
\item The time evolution of the initial state is under control as long as we are in the regime $|{\vec p}|/\mu\ll 1$, and $\hat{c}_Tt\mu\ll 1$.  In particular, we do not expect the theory to remain perturbative for arbitrarily long times, when the wavefunction has completely delocalized. (Indeed, even the classical theory becomes non-linear at late times, due to the growth of classical vortex perturbations.).
\end{itemize}

 \section{Results for the stress tensor response}
 \label{sec:results}
\begin{figure}[t]
\centering
\includegraphics[scale=0.25]{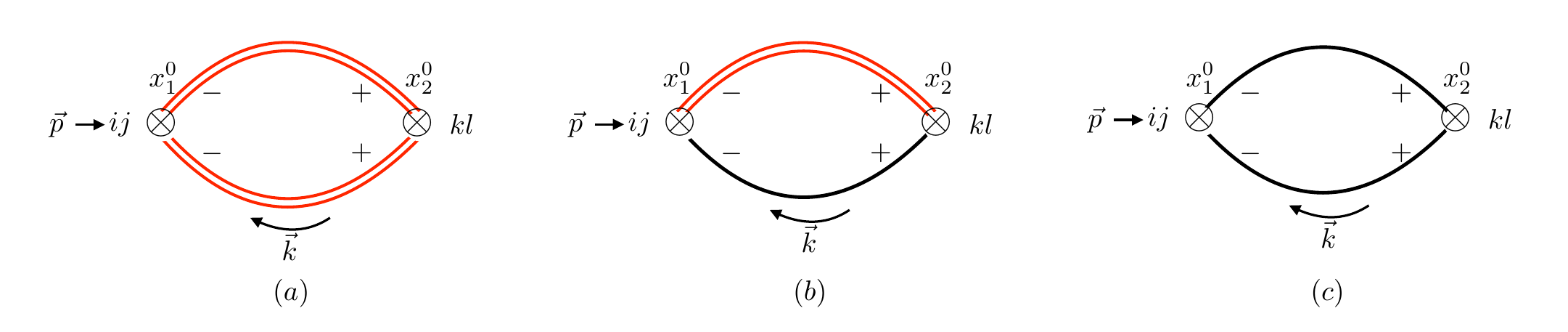}
\caption{Leading order Feynman diagram contributions to the momentum space connected Wightman function $\langle T_{ij}(x_1) T_{kl}(x_2)\rangle=\langle T^-_{ij}(x_1) T^+_{kl}(x_2)\rangle$ from (a) $TT$  (b) mixed $TL$, (c) and $LL$ two-phonon (``superfluid'') internal propagators.} 
\label{fig:2pt}
\end{figure}

\noindent  For simplicity, we restrict ourselves to the response function of the trace-free part of the stress tensor $[T_{ij}]^{TF}=T_{ij}-{1\over d} \delta_{ij} T^k{}_k$, which starts out at quadratic order in the fluctuations, 
  \be
  [T^{ij}]^{TF}= w_0 [\partial_t \pi^i\partial_t \pi^j]^{TF}+\cdots.   
  \ee
  To leading order in the power counting, the momentum space\footnote{By convention, in the spatially translation invariant state $|\psi_i\rangle$,  a generic momentum-space Green's function of local operators $A(x_1),$ $B(x_2)$ is denoted $G^{AB}_{\vec p}(x^0_1,x^0_2) = \int d^d {\vec x} e^{i {\vec p}\cdot{\vec x}_{12}} A(x_1) B(x_2),$ ${\vec x}_{12}={\vec x}_1-{\vec x}_2$.}  connected Wightman two-point function for this operator is given by the one-loop Feynman diagrams in Fig.~\ref{fig:2pt}, where we explicitly separate into the contribution from two virtual $T$ modes, ($\int_{\vec k}\equiv\int {d^d{\vec k}\over (2\pi)^d}$)
\be
\label{eq:TT}
TT={{\hat c}_T^2\over {4\mu^{2\Delta-2}}}\left[\int_{\vec k}{P_{ik}^T({\vec k}) P_{jl}^T({\vec k}+{\vec p})\over |{\vec k}|^{-\Delta} |{\vec k}+{\vec p}|^{-\Delta}}+\mbox{perms}\right]^{TF},
\ee 
a mixed term involving one longitudinal phonon and one vortex propagator,
\be
\label{eq:TL}
TL={{\hat c}_T c_s\over {4\mu^{\Delta-1}}}\left[\int_{\vec k} e^{-i c_s t |{\vec k}|}{P_{ik}^L({\vec k}) P_{jl}^T({\vec k}+{\vec p})\over |{\vec k}|^{-1} |{\vec k}+{\vec p}|^{-\Delta}}+\mbox{perms}\right]^{TF},
\ee
where $t=x^0_1-x^0_2$, and a `superfluid' contribution from the two $L$-phonon intermediate state
\be
\label{eq:LL}
LL={c_s^2\over 4}\left[\int_{\vec k}  e^{-i c_s t \left(|{\vec k}| +|{\vec k}+{\vec p}|\right)}{P_{ik}^L({\vec k}) P_{jl}^T({\vec k}+{\vec p})\over |{\vec k}|^{-1} |{\vec k}+{\vec p}|^{-1}}+\mbox{perms}\right]^{TF}.
\ee

\noindent By rotational invariance, these integrals can be expressed as linear combinations of a tensor basis constructed from $\delta^{ij}$ and the external momentum ${\vec p}$, with scalar coefficients that can be reduced to a set of master Feynman integrals.   For the cases $TT, TL$ the master integrals are defined by dimensional regularization in $d$-spatial dimensions and can be equivalently evaluated by the method of regions~\cite{Beneke:1997zp} or by the Mellin-Barnes method~\cite{Smirnov:2006ry}, see Appendix~\ref{app:master}.   In dimensional regularization, the relevant integrals are free of UV or IR divergences for generic values of the spectral index $\Delta$, in either $d=2,3$ spatial dimensions, as long as we define the complex phase in Eq.~(\ref{eq:TL}) via the prescription 
\be
t\rightarrow t - i 0^+.
\ee

Using this same $i\epsilon$ prescription for the phases in Eq.~(\ref{eq:LL}), the $LL$ term is manifestly free of UV or IR divergences, and can be directly evaluated in $d=2,3$ without the need for regularization.  While it may be possible to evaluate Eq.~(\ref{eq:LL}) in arbitrary $d$ in terms of generalized hypergeometric functions (via their Mellin-Barnes integral representation), we only considered the case $d=3$ where it reduces to elementary functions, see Eq.~(\ref{eq:LLresult}) below.

Of the various tensor structures that arise in these integrals, we will focus on the scalar coefficient function multiplying the lowest harmonic $Y_0^{ij,kl} ={1\over 2}\left(\delta^{ik} \delta^{jl}+\delta^{il} \delta^{j}-{2\over d} \delta^{ij} \delta^{kl}\right)$ appearing in  the correlator. In $d=3$,   the $TT$ and $LL$ contributions to the coefficient of $Y_0^{ij,kl}$ in the Wightman function are expressible in terms of elementary functions of ${\vec p}{\,}^2$ and $t$, while $TL$ involves both elementary functions as well as ${}_1 F_2$ and ${}_3 F_2$ hypergeometric functions.   However, it turns out that for the case $d=3$, these special functions drop out from $G^R_{\vec p}(t)$, defined as the $Y_0^{ij,kl}$ coefficient function of the \emph{retarded} correlators.  Because (in any spatial dimension) the pure $TT$ contribution is time independent at one-loop, its contribution  to the response function vanishes altogether.    Therefore, although the causal response function is trivial in the incompressible limit $c_s\rightarrow\infty$ in which the $L$-phonons decouple, the result for $G^R_{\vec p}(t)$ at finite $c_s$ differs from that of a pure superfluid.

\begin{figure}[t]
\centering
\includegraphics[scale=0.25]{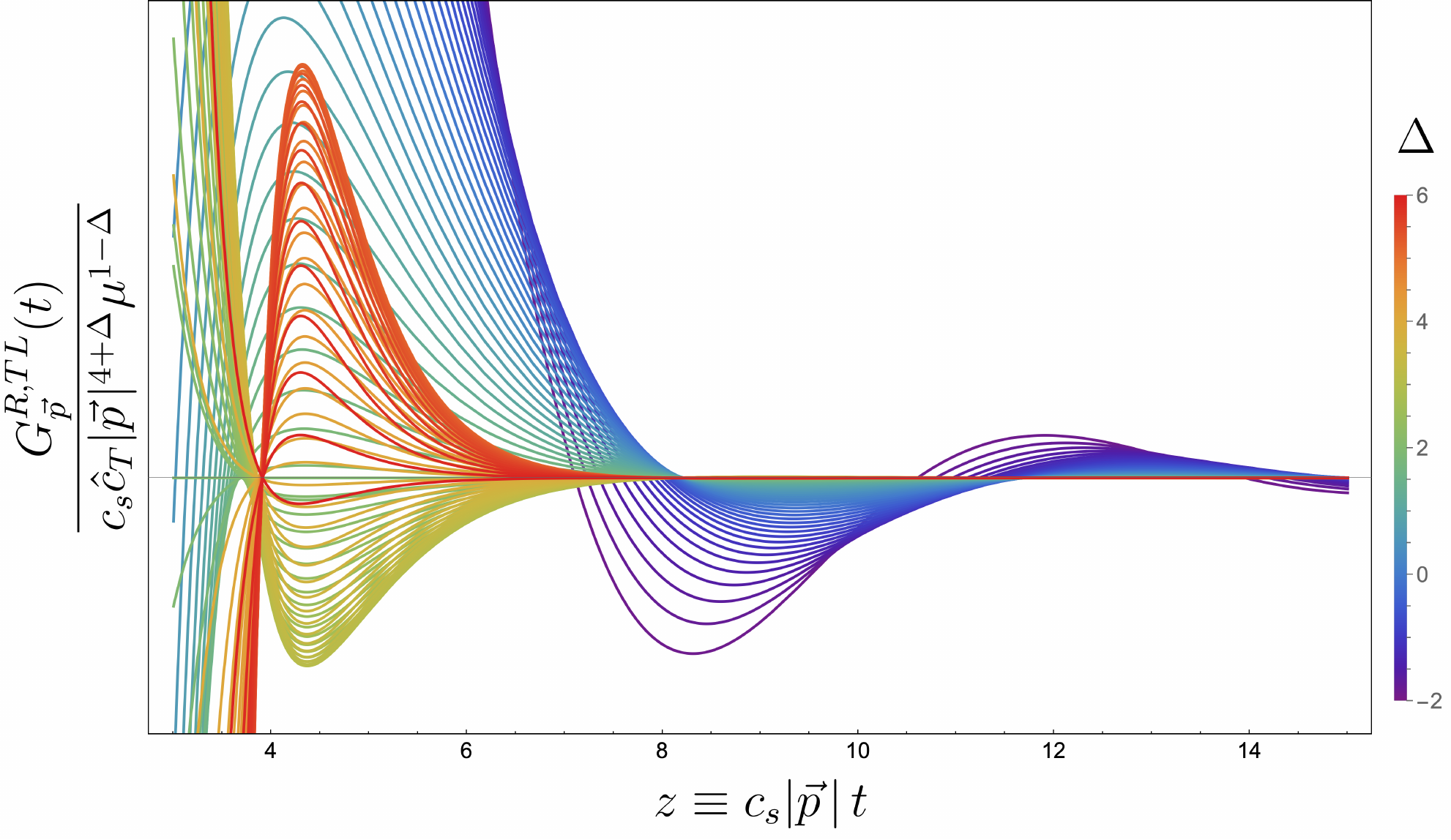}
\caption{$TL$ contribution to the response function, given by Eq.~\eqref{eq:LTresult}, plotted for a range of values of the spectral parameter $\Delta$.} 
\label{fig:LT}
\end{figure}

We have $G^R_{\vec p}(t)=G^{R,TL}_{\vec p}(t)+G^{R,LL}_{\vec p}(t),$ where in $d=3$, the mixed term in Fig.~\ref{fig:2pt}(b) is 
\bea\label{eq:LTresult}
\nonumber
G^{R,TL}_{\vec p}(t) &=&{\hat c}_T\frac{(\Delta+3) \Gamma (\Delta+2)}{\Delta }\sin\left({\pi\Delta \over 2}\right)\\
& &{}\times  \left(|{\vec p}|\over\mu\right)^{\Delta-1} \theta(t) {g^{TL}_R(c_s t |{\vec p}|)\over\pi^2 c_s^4 t^5},
\eea
\begin{widetext}
\bea
\nonumber
g^{TL}_R(z) = \left[{(\Delta+4)(\Delta+5) z^2-6(\Delta+5) (\Delta+7)\over z^{\Delta+{3}}}\right]\cos z + \left[{(\Delta+2) z^4 - 3 (\Delta+5) (\Delta+6) z^2 + 6(\Delta+5) (\Delta+7)\over z^{\Delta+4}}\right] \sin z,
\eea
\end{widetext}
for $z>0$.  The two-phonon vacuum polarization, Fig.~\ref{fig:2pt}(c), is given by
\be\label{eq:LLresult}
G^{R,LL}_{\vec p}(t)=\theta(t) {g^{LL}_R(c_s t |{\vec p}|)\over 60 \pi^2 c_s^3 t^5},
\ee
with $g^{R,LL}(z) = (z^2 -3) \cos z - 3 z \sin z$.   
This last term in particular describes the response of a superfluid in its finite energy density ground state (phonon vacuum), so that $G^{R,TL}_{\vec p}(t)$, plotted in Fig.~\ref{fig:LT} for varying parameter $\Delta$, encodes deviations from superfluidity resulting from the time evolution of the vortex wavefunction.

\section{Discussion and conclusions}
\label{sec:conc}

\noindent In this paper, we have studied the quantum mechanical time evolution of perfect fluids in semi-classical initial states of the form Eq.~(\ref{eq:state}).    We focused on the stress tensor response function in this class of states, finding results that, at least in certain kinematic regimes, are calculable in perturbation theory without the need to introduce any ad hoc parameters to the Lagrangian in order to lift the infinite degeneracy of vortex modes implied by classical SDiff invariance.  The resulting prediction for the retarded correlator becomes trivial in the limit of infinite compressibility $(c_s\rightarrow\infty)$ in which the phonons decouple, but differs from the superfluid response at finite $c_s$.

In our analysis, we limited ourselves to studying the leading order quantum corrections in one specific channel.  However, we believe that our conclusions apply more broadly, e.g., to correlators of generic local operators in this theory, regardless of whether they are SDiff singlets, and even after incorporating the effects of cubic and higher self-interactions, which are controlled by powers of $1/w_0$ in the derivative expansion.

Besides explicitly verifying that loop effects due to interactions are under analytical control, there are still a number of open conceptual and practical questions to address.  On the practical side, it would be interesting to compute observables in more general states, possibly ones that break parity $\mathcal{P}$ and time-reversal $\mathcal{T}$ explicitly at $t=0$, for instance by having a $t=0$ velocity profile $\langle{\vec v}({\vec x})\rangle\neq 0$, as well as states that contain non-Gaussianities at $t=0$.  For states that break $\mathcal{T}$, the one-point correlators $\langle T^{ij}(x)\rangle$ and $\langle {\vec v}(x)\rangle$ are in general non-zero at time $x^0>0$, allowing direct comparison with the transport properties of a classical fluid. 

 More conceptually, it would be interesting to make contact with the picture of the quantum perfect fluid developed in~\cite{Dersy:2022kjd,Cuomo:2024ekf}, which describes the physics near the true non-perturbative vacuum of the theory.  For example, is the Clebsch representation of the fluid employed in~\cite{Dersy:2022kjd,Cuomo:2024ekf} quantum mechanically equivalent to the dynamics as formulated in this paper, by path integral quantization of the Lagrange variable $\phi^I(x)$?   If so, can we find a description of the vorton quasiparticle excitations predicted by these authors in terms of Schrodinger functional states $\Psi[\phi^I]$ of the Lagrange space field?   We hope to address some of these problems in future work.

\section*{Acknowledgments}
\noindent We thank the organizers and participants of the EPFL Bernoulli Center workshop \emph{Effective Field Theories for Hydrodynamics} for their hospitality and for lively discussions.   W.D.G. was supported by the U.S. Department of Energy Grant No. DE-SC00-17660.   The work of P.T. has been supported in part through a joint United Kingdom Research and Innovation (UKRI) partnership with the U.S. National Science Foundation (NSF) under Grant No. EP/Z003423/1.

\appendix
\section{IR regularization scheme dependence of SDiff invariant Green's functions}\label{app:triumph}

\noindent In their search for well-defined observables that are not sensitive to the infinite degeneracy of the zero energy vortex modes, ref.~\cite{Endlich:2010hf} deformed the classical fluid Lagrangian by the addition of a term 
\begin{equation}
\label{eq:DeltaS}
\Delta {\cal L} = -\frac{w_0}{2} c_T^2 B^{II}
\end{equation}
which breaks explicitly the symmetry under volume preserving diffeomorphisms acting on $\phi^I$.    For $c_T\neq 0,$ the transverse mode dispersion relation becomes  $\omega_T({\vec p}) =  c_T |\vec{p}|$, and the quantum theory has a normalizable vacuum state as well as a tower of multi-particle $L,T$-mode Fock states.  Thus, at $c_T\neq 0$, it is possible to compute $S$-matrix elements involving these asymptotic states and their corresponding cross sections and decay rates by summing over phase space in the usual manner.   

While it is possible to find isolated examples of phonon-vortex processes with a well-behaved $c_T\rightarrow 0$ limit, generically one finds that already at tree-level, cross sections have power divergences as the regulator in Eq.~(\ref{eq:DeltaS}) is removed.   For example the rate for phonon decay $L\rightarrow TT$ diverges as $1/c_T^3$ if the energy of the initial phonon is held fixed to a finite value~\cite{Endlich:2010hf} .   This singularity simply reflects the fact that the final state particles carry momenta of order $|{\vec p}_T|\sim E_L/c_T$, so the phase space volume diverges as $c_T\rightarrow 0$.   Such divergences become more severe as more $T$-mode particles are added to the final state, making it difficult to identify observables that become ``IR safe'' (independent of the regulator) by the familiar procedure of constructing inclusive observables where one sums over all possible finite states below some soft energy scale set by the  resolution of the experiment.

The failure to construct a perturbative $S$-matrix that remains well-defined for $c_T\rightarrow 0$ does not necessarily mean that the quantum theory lacks well-defined observables.  A logical possibility is that quantum mechanically, the SDiff invariance of the classical action becomes \emph{gauged}, in the sense that both the true physical states and the observables are defined to be invariant under SDiffs.   Under this assumption, an experimentalist would not be able to prepare states or measure operators that are in non-trivial representations of the SDiff algebra.   Only the invariant Green's functions need be unambiguously calculable, up to local counterterms that are invariant.

Ref.~\cite{Gripaios:2014yha} calculated one-loop examples of such invariant (time-ordered) two-point functions, for a fluid in $d=2$,  finding results that are consistent with the gauge symmetry realization of the SDiff symmetry.  The authors defined the relevant Green's functions directly in terms of the momentum space Feynman rules implied by Eq.~(\ref{eq:action}), expanded in powers of the fluctuation $\pi^i(x)$ about the classical ground state.   For example, the time-ordered propagator has a $T$-mode pole of the form $i P^{ij}_T({\vec p})/(p_0^2+i\epsilon)$, whose IR divergent contributions inside loop integrals need to be suitably regularized.    To do so, ref.~\cite{Gripaios:2014yha} employed a form of dimensional regularization in which the energy and the momentum integrals are \emph{independently} continued to non-integer dimensions.  In a number of illustrative cases, this algorithm was found to yield SDiff invariant two-point functions which are free of IR divergences as the regulator is removed, with non-local contributions that arise from all three $TT, TL, LL$ channels.

Because in dimensional regularization power (UV or IR) divergent contributions are defined to be identically zero, it is not clear how the individual potentially IR divergent Feynman diagrams that contribute to an SDiff invariant Green's function combine to yield results that are free of singularities.   In order to test the dependence on the choice of regularization scheme, we instead computed SDiff invariant correlators using Eq.~(\ref{eq:DeltaS}) as regulator.  In this scheme, each individual diagram is a polynomial in powers of  $1/c_T$, and one can explicitly check whether the various terms that contribute to a given correlation function add up to an expression that is independent of $c_T$ before taking the limit $c_T\rightarrow 0$.  Working at finite $c_T$, we calculated the connected Wightman density-density correlator $\langle\delta\rho \delta\rho\rangle$ for a fluid in $d=2,3$ (still using dimensional regularization for the spatial loop momenta) and the connected Wightman function $\langle [T_{ij}]^{TF} [T_{kl}]^{TF}\rangle$ of the trace-free stress tensor.

In all cases we computed, we obtained results that contain power singularities in the $c_T\rightarrow 0$ limit, suggesting that  IR safety of gauge invariant observables is not guaranteed, but depends crucially on the method used to regularize perturbation theory.   The simplest example is that of the stress tensor two-point function, which in the incompressible limit $c_s\rightarrow \infty$ is given at leading order by a one-loop Feynman diagram, of the same topology as Fig.~\ref{fig:2pt}(a).  Because only a single Feynman diagram contributes for $c_s\rightarrow \infty$, IR safety of the observable would require the loop integral to scale at worse as $c_T^0$ in the $c_T\rightarrow 0$ limit.  

At finite $c_T$, the regulator guarantees that the theory has a normalizable vacuum state and a non-degenerate free field Wightman function, which in momentum space $p^\mu=(p^0,{\vec p})$ takes the form
\be
W^{ij}(p) =  w_0^{-1} P^{ij}_T({\vec p})\delta_+\left(p^\mu\right),
\ee
with $\delta_{\pm}(p^\mu)\equiv 2\pi\theta(\pm p^0) \delta(p_0^2-c_T^2{\vec p}^2)$.   The one-loop contribution to $\langle [T_{ij}]^{TF} [T_{kl}]^{TF}\rangle$ is then 
\begin{widetext}
\be
TT=\int{d^d{\vec k}\over (2\pi)^d}\int_{-\infty}^{\infty}{dk^0\over 2\pi} \delta_-(k^\mu) \delta_+(p^\mu+k^\mu) k_0^2 (k_0+p_0)^2 \left[P^{ik}_T({\vec k}) P^{jl}_T({\vec k}+{\vec p})+\mbox{perms}\right]^{TF},
\ee
\end{widetext}
where the external momentum is $p^\mu=(p^0,{\vec p})$ and the factor $k_0^2 (k_0+p_0)^2$ reflects the presence of time derivatives in the ${\cal O}({\vec\pi}^2)$ part of $[T_{ij}]^{TF}$.  

The result for the $k^0$ integral depends on the order in which the limit $c_T\rightarrow 0$ is taken.    If we take $c_T=0$ first, before calculating any integrals, the expression factorizes into the product of a spatial momentum integral and a $k^0$ integral of the form
\begin{widetext}
\be
\label{eq:I0}
I(p_0)=\int_{-\infty}^{\infty}{dk^0\over 2\pi} \theta(- k^0)  \theta(k^0+p^0)k_0^2 (k_0+p_0)^2 \delta(k_0^2) \delta((k_0+p_0)^2)
\ee
\end{widetext}
whose value depends on the definition of the singular distribution\footnote{For the Feynman (time-ordered) two-point function, one instead inserts the Feynman propagator ${i w_0^{-1}P^{ij}_T({\vec p})/(p_0^2-c_T^2 {\vec p}^2+i0^+)}$ inside the loop integral.    Taking $c_T=0$ before evaluating the integral, the result is in this case proportional to the UV divergent integral $\int {dk^0\over 2\pi} \cdot 1$.  Regularizing the $k^0$ integral by dimensional continuation of the time coordinate to $d_t=1-\epsilon_t$ dimensions, $\epsilon_t\rightarrow 0$  would then simply define this scaleless integral to vanish even, before taking the physical limit $d_t\rightarrow 1$.} $\delta(k_0^2)$, but can be taken to be $I(p_0)\equiv 0$ since it is being integrated against a function that vanishes sufficiently rapidly at $k_0^2$.

On the other hand, performing the $k^0$ integral at finite $c_T$ first yields a spatial momentum integral 
\begin{widetext}
\be
TT = {\pi\over 2} c_T^2 \int{d^d{\vec k}\over (2\pi)^d}  \theta(p_0-c_T|{\vec k}|) \delta(p_0-c_T|{\vec k}|-c_T|{\vec k} +{\vec p}|) \left[{P^{ik}_T({\vec k})  P^{jl}_T({\vec k}+{\vec p}) \over |{\vec k}|^{-1}|{\vec k}+{\vec p}|^{-1}}+\mbox{perms}\right]^{TF}.
\ee
\end{widetext}
This is of the form of a two-particle phase space integral over on-shell vortex states with total energy $p_0$.  For fixed $p_0$, each particle has momentum of order $p_0/c_T$, so the integral scales as $1/c_T^{d+2}$ in the $c_T\rightarrow 0$ limit.  

Indeed, by an explicit calculation we have found that the coefficient function of the $Y^{ij,kl}_0$ tensor structure appearing in the $d=3$ retarded Green's function is given by
\be
\label{eq:TTcTsing}
G^{R,TT}_{\vec p}(t)=-\theta(t){c_T^2\over 40\pi^2}\left[ {7\over (c_T t)^5} + {{\vec p}{\,}^2\over 2(c_T t)^3}\right]+{\cal O}(c_T),
\ee 
in the $(t,{\vec p})$ mixed representation.  The same $1/c_T^3$ singularity implied by power counting now shows up as non-local in time but spatially local response of the fluid.  We have also verified that away from the incompressible limit, the $TL$ contribution of Fig.~\ref{fig:2pt}(b) depends linearly on $c_T$, whereas the $LL$ term is obviously finite and independent of the regulator.  

The  $\langle [T_{ij}]^{TF} [T_{kl}]^{TF}\rangle$ channel is technically the simplest to calculate one-loop observable in this theory.  However, it is not the most illustrative example from the point of view of showing how non-local features of SDiff invariant Green's functions depend on the precise regularization procedure used to compute them.  For example, we have checked that in the $\langle \delta\rho\delta \rho\rangle$ channel, the results also depend crucially on the order of limits.  Evaluating the relevant integrals before taking the $c_T\rightarrow 0$ limit yields a response function that in $d=2,3$ is non-local in both space and time, with a singular coefficient that scales as $1/c_T^3$ in $d=3$ (arising from unitarity cuts related to the $L\rightarrow TT$ decay rate, $\Gamma(L\rightarrow TT)\sim 1/c_T^3$).   On the other hand, taking $c_T=0$ before evaluating any integrals generates expressions that are undefined even after analytic continuation in the number of time dimensions.   For example, the analog of Eq.~(\ref{eq:I0}) in the $\langle\delta\rho\delta\rho\rangle$ connected Wightman function does not have the energy factors $k_0^2 (k_0+p_0)^2$, so that the integral over $\delta(k_0^2)$ is no longer meaningful.

The analysis in this appendix shows that different prescriptions for regularizing the vortex modes yield physically inequivalent predictions for the fluid's response.  In light of this,  it seems that the assumption that the SDiff's are gauge symmetries is not sufficient by itself to provide a robust solution to the vortex degeneracy problem of the quantum perfect fluid.

\section{Master integrals}\label{app:master}
\noindent To obtain the results of Sec.~\ref{sec:results}, we reduced the integrals in Figs.~(\ref{fig:2pt})(a),(b) to linear combinations of master integrals of the form 
\be
I_{\alpha,\beta}({\vec p},z)\equiv\int_{\vec k} {e^{-z|{\vec k}|}\over |{\vec k}|^{2\alpha} |{\vec k}+{\vec p}|^{2\beta}},
\ee
where $z=0$ for the $TT$ term in Fig.~(\ref{fig:2pt})(a) and $z=-i c_s t +0^+$ in Fig.~(\ref{fig:2pt})(b).   We have been able to obtain these master integrals using the Mellin-Barnes method, by inserting the Mellin transform of the exponential function
\be
e^{-z}=\oint_\Gamma {ds\over 2 \pi i} z^{-s} \Gamma(s),
\ee
where the closed contour $\Gamma$ runs along the $\mbox{Im} s$ axis, avoiding the pole at $s=0$.  This identity then allows us to evaluate the integral over ${\vec k}$ using the standard result for one-loop (massless) self-energy diagrams in dimensional regularization, i.e.
\bea
\nonumber
I_{\alpha,\beta}({\vec p},0) &=& {\Gamma\left({d\over 2}-\alpha\right)\Gamma\left({d\over 2}-\beta\right)\over \Gamma(\alpha)\Gamma(\beta)}  {\Gamma\left(\alpha+\beta-{d\over 2}\right)\over \Gamma(d-\alpha-\beta)}\\
& & {} \hspace{0.5cm}\times  {|{\vec p}|^{d-{\alpha+\beta\over 2}}\over (4\pi)^{d/2}},
\eea
so that, for $z\neq 0$,
\be
I_{\alpha,\beta}({\vec p},z) = {\Gamma\left({d\over 2}-\beta\right)\over\Gamma(\beta)}{|{\vec p}|^{d-{\alpha+\beta\over2}}\over (4\pi)^{d/2}} f_d(z|{\vec p}|),
\ee
with 
\be
f_d(z) = \oint_\Gamma {ds\over 2\pi i}  z^{-s} {\Gamma(s) \Gamma\left({d+s\over 2}-\alpha\right) \Gamma\left(\alpha+\beta+{s-d\over 2}\right)\over\Gamma\left(\alpha-{s\over 2}\right)\Gamma\left(d+{s\over 2}-\alpha-\beta\right)}.
\ee

For generic values of $d,\alpha,\beta$, $f_d(z)$ is a linear combination of ${}_3 F_2$ hypergeometric functions of $z^2/4$ with coefficients that depend on powers of the variable $z$ itself.   However, the full expression is not needed in order to obtain the results described in Sec.~\ref{sec:results}.   For $d=3$ and the specific values of $1-2\alpha=2m\geq 0$  that arise in our calculation, the combination $g_3(z)=f_3(z)-f_3(-z)$ needed to calculate the retarded Green's functions can be expressed in terms of elementary functions of $z$,
\bea
\nonumber
g_3(z)&=&-{2\over\sin\pi\beta} \left[\left({z\over 2}\right)^{2\alpha+2\beta-3} -\left({-z\over 2}\right)^{2\alpha+2\beta-3}\right]\\
& & {} \hspace{1cm}\times p_\alpha\left(z{d\over dz}\right) {\sinh z\over z},
\eea
where $p_\alpha(x)$ is the following polynomial in $x$,
\be
p_\alpha(x) =2^{2\alpha-1} {\Gamma(3-2\beta+ x-2\alpha)\over\Gamma(2-2\beta + x)}
\ee
defined for $1-2\alpha=2m$, with $m\geq 0$ a non-negative integer.

\bibliographystyle{apsrev4-2}
\bibliography{references}{}

\end{document}